\documentclass[a4paper,12pt]{article}
\pdfoutput=1
\usepackage{cite}
\usepackage{amsmath}
\usepackage{amsfonts}
\usepackage{amssymb}
\usepackage{graphicx, rotating}
\usepackage{epsfig}
\usepackage{latexsym}
\usepackage{graphicx}
\usepackage{color}
\usepackage{amsmath,bm,amssymb}

\usepackage{color}
\usepackage[english]{babel}
\usepackage{hyperref}

\newcommand{\Slash}[1]{{\ooalign{\hfil#1\hfil\crcr\raise.167ex\hbox{/}}}}

\newcommand{\beq}{\begin{equation}}  \newcommand{\eeq}{\end{equation}}
\newcommand{\bef}{\begin{figure}}  \newcommand{\eef}{\end{figure}}
\newcommand{\bec}{\begin{center}}  \newcommand{\eec}{\end{center}}
\newcommand{\non}{\nonumber}  
\newcommand{\laq}[1]{\label{eq:#1}}  

\newcommand{\Eq}[1]{Eq.~(\ref{eq:#1})}
\newcommand{\Eqs}[1]{Eqs.~(\ref{eq:#1})}
\newcommand{\eq}[1]{(\ref{eq:#1})}

\newcommand{\ab}[1]{\left|{#1}\right|}
\newcommand{\vev}[1]{ \left\langle {#1} \right\rangle }

\newcommand{\SU}[1]{{\rm SU{#1} } }

\def\({\left(}
\def\){\right)}

\def\O{\mathcal{O}}
\def\U{\mathop{\rm U}}

\newcommand{\AND}{~{\rm and}~}
\newcommand{\EV}{ {\rm \, eV} }

\newcommand{\MEV}{ {\rm \, MeV} }
\newcommand{\GEV}{ {\rm \, GeV} }
\newcommand{\TEV}{ {\rm \, TeV} }

\def\a{\alpha}

\def\d{\delta}

\def\f{\phi}

\def\h{\theta}
\def\k{\kappa}
\def\l{\lambda}

\def\x{\xi}

\def\D{\Delta}

\def\H{\Theta}
\def\L{\Lambda}

\def\tl{\tilde}
\def\*{\dagger}

\begin{document}
\renewcommand\bibname{\Large References}

\begin{center}

\hfill   TU-1091\\
\hfill  IPMU19-0112\\

\vspace{1.5cm}

{\Large\bf  QCD Axion on Hilltop by a Phase Shift of $\pi$ }\\
\vspace{1.5cm}

{\bf  Fuminobu Takahashi$^{1,2}$, Wen Yin$^{3}$}

\vspace{12pt}
\vspace{1.5cm}
{\em 
$^{1}$Department of Physics, Tohoku University,  
Sendai, Miyagi 980-8578, Japan \\
$^{2}$Kavli Institute for the Physics and Mathematics of the Universe (WPI),
University of Tokyo, Kashiwa 277--8583, Japan\\
$^{3}${ Department of Physics, KAIST, Daejeon 34141, Korea} \vspace{5pt}}

\vspace{1.5cm}
\abstract{
We show that the initial misalignment angle of the QCD axion (or axion-like particles) can be set 
very close to $\pi$,  
if the QCD axion has a mixing with another heavy axion which induces the phase shift $\approx \pi$ after inflation.
In the simplest case, the heavy axion plays the role of the inflaton, and we call such inflation as ``$\pi$\hspace{-0.2mm}nflation."
The basic idea was first proposed by Daido and the present authors in Ref.~\cite{Daido:2017wwb} 
in 2017 and more recently discussed in Ref.~\cite{Takahashi:2019qmh}. We show that 
the QCD axion with a decay constant $f_a \gtrsim 3 \times 10^9$\,GeV
 can explain dark matter by the $\pi$\hspace{-0.2mm}nflation mechanism.
A large fraction of the parameter region has an overlap with the projected sensitivity of 
ORGAN, MADMAX, TOORAD and IAXO. We also study implications for the effective neutrino species and isocurvature perturbations.
The $\pi$\hspace{-0.2mm}nflation can
provide an initial condition for the hilltop inflation in the axion landscape, and in a certain set-up,
a chain of the hilltop inflation may take place.  
}

\end{center}
\clearpage

\setcounter{page}{1}
\setcounter{footnote}{0}

\section{Introduction}

The QCD axion \cite{Peccei:1977hh,Peccei:1977ur,Weinberg:1977ma,Wilczek:1977pj} is a plausible candidate for dark matter (DM). It starts to oscillate about the CP conserving minimum, when its temperature-dependent mass becomes comparable to the Hubble parameter during the QCD phase transition~\cite{Preskill:1982cy,Abbott:1982af,Dine:1982ah}. 
The abundance of the QCD axion generated by the misalignment mechanism is given by~\cite{Bae:2008ue, Visinelli:2009zm,Ballesteros:2016xej}
\beq
\laq{ab}
\Omega_a h^2 
\,\simeq\, 
0.0092 F(\h_i )\h_i^{2}
\left(\frac{f_a}{10^{11}\,{\rm GeV}}\right)^{1.17}, 
\eeq
where $$\theta_i\equiv \frac{a_i}{f_a}$$ is the initial misalignment angle, and
$f_a$ is the axion decay constant. The coefficient $F(\theta_i)$ is given by
\beq
F(\theta_i)=\left[\log{\(\frac{e}{1-\frac{\theta_i^2}{\pi^2}}\)}\right]^{1.17},
\eeq
which takes account of the anharmonic effect. Here and in what follows the QCD axion 
is stabilized at the CP conserving minimum, $a = 0$, in the present vacuum.

\begin{figure}[!t]
\begin{center}  
   \includegraphics[width=105mm]{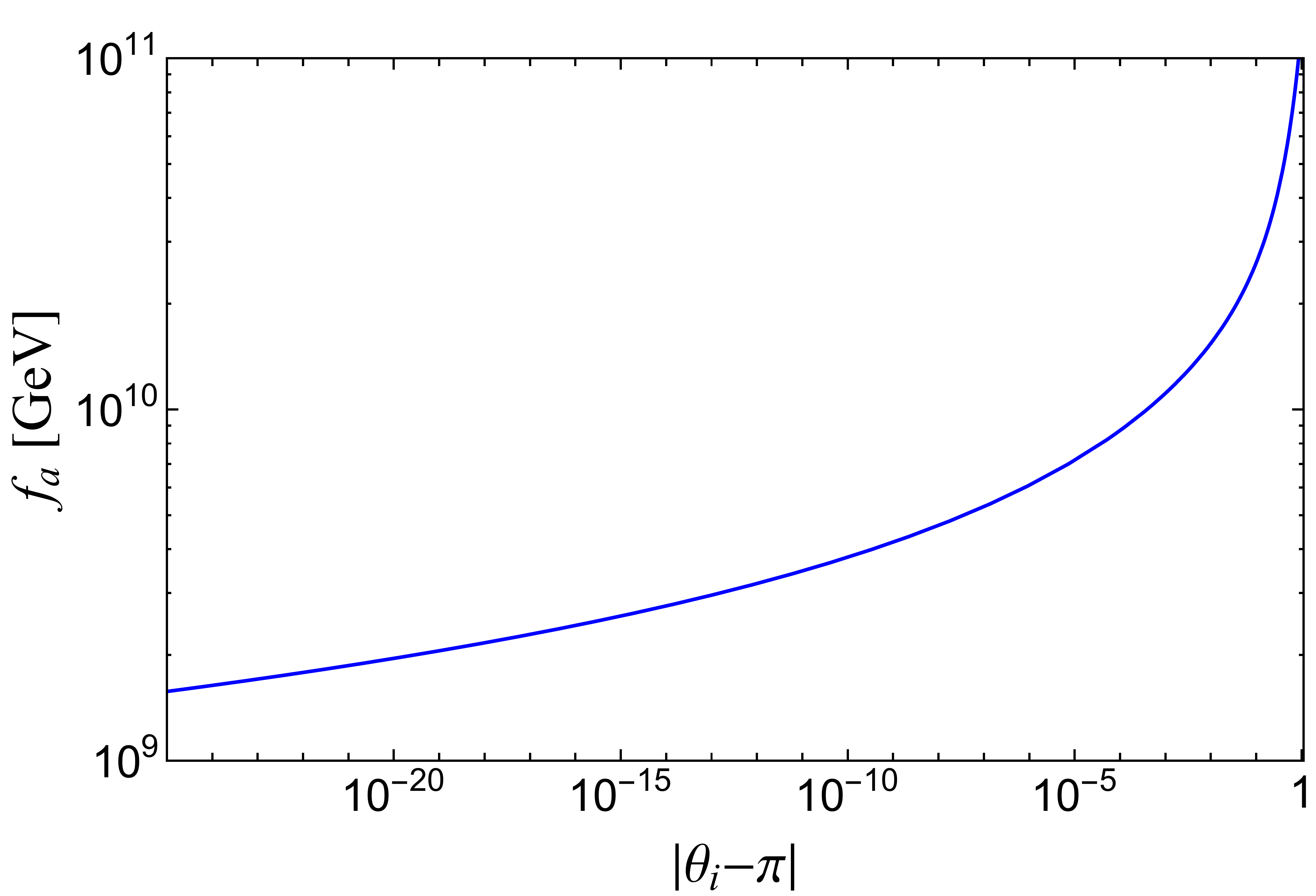}
      \end{center}
\caption{
The  relation between $f_a$ and $|\theta_i-\pi|$ explaining the DM abundance $\Omega_a  h^2=0.12.$ }\label{fig:DM} 
\end{figure}

One can see from \Eq{ab} that the QCD axion 
explains the observed DM abundance, $\Omega_{\rm DM}h^2\simeq 0.12$~\cite{Aghanim:2018eyx}, 
for $\theta_i = {\cal O}(1)$ and $f_a \simeq 10^{12}$\,GeV. This sets
the upper bound of the so-called classical axion window, $f_a \lesssim 10^{12}$\,GeV.  This does not preclude the possibility of larger or smaller values of $f_a$. Larger $f_a$ is possible if $\theta_i$ is (much) smaller than unity. This can be realized by fine-tuning based on the anthropic argument~\cite{Linde:1991km,Wilczek:2004cr,Tegmark:2005dy}, or very low-scale inflation~\cite{Graham:2018jyp,Guth:2018hsa}.\footnote{Alternatively, one may modify thermal history of the Universe~\cite{Dine:1982ah,Steinhardt:1983ia,Lazarides:1990xp,Kawasaki:1995vt,Kawasaki:2004rx}, or introduce the explicit PQ breakig by the Witten effect to suppress the QCD axion abundance~\cite{Witten:1979ey,Kawasaki:2015lpf,Nomura:2015xil,Kawasaki:2017xwt}.}  
On the other hand, smaller $f_a$ is possible if 
 the initial position of the QCD axion is close to the hilltop of the potential, i.e.,
$\theta_i \approx \pi$.
In the hilltop limit, the anharmonic coefficient, $F(\theta_i)$, logarithmically increases; e.g. $F(\pi-10^{-5})\simeq 20$  and  $F(\pi-10^{-10})\simeq 42.$  
As a result, the QCD axion abundance can be enhanced if $\theta_i$ is close to $\pi.$
In Fig.\,\ref{fig:DM} we show the relation between $f_a$ and $|\theta_i-\pi|$
to explain the observed DM abundance.

In the hilltop limit, the power spectrum of the axion isocurvature perturbation~\cite{Lyth:1991ub,Kobayashi:2013nva} as well as its non-Gaussianity~\cite{Kobayashi:2013nva} are also 
significantly enhanced.\footnote{
Before Ref.\,\cite{Kobayashi:2013nva}, 
there were some inconsistencies among the literature concerning the power spectrum of the axion isocurvature perturbations
in the hilltop limit.}  Therefore, the axion DM with relatively small $f_a$ poses two issues. One is the fine-tuning of the initial misalignment angle near $\pi$. The other is that one needs very low-scale inflation to satisfy the isocurvature bound.

Smaller $f_a$ implies a heavier QCD axion mass, which is challenging  from the experimental point of view. Many experiments for axion DM search have been proposed so far, and some of them are aiming at such relatively heavy axion masses. The range of the axion mass partially overlaps with the  mass range expected by the axion production mechanism using the string-wall network~\cite{Klaer:2017ond,Gorghetto:2018myk,Kawasaki:2018bzv,Buschmann:2019icd,Hindmarsh:2019csc}. 
However, there are currently uncertainly in the estimate as one has to rely on a large amount of extrapolation 
from the parameters used in the numerical calculations
to the realistic ones~\cite{talk}. 

In this paper we provide a mechanism to set the initial position of the QCD axion very close to the hilltop of the potential. We consider an inflation model where another axion plays the role of the inflaton. Being the axion, the inflaton has a  potential with the periodicity $2\pi f_\phi$, where $f_\phi$ is the inflaton decay constant. For instance, successful slow-roll inflation is possible if the potential consists of two cosine terms, which is known as the multi-natural inflation~\cite{Czerny:2014wza, Czerny:2014xja,Czerny:2014qqa,Higaki:2014sja, Croon:2014dma}\footnote{
See Ref.~\cite{Daido:2017wwb, Daido:2017tbr, Takahashi:2019qmh} for realization of the multi-natural inflation
in terms of an axion-like particle.}. The inflation takes place on the flat plateau around 
the potential maximum, and after inflation ends, the inflaton rolls down toward the nearest potential minimum. The distance between the maximum and the minimum is naturally equal to or very close to $\pi$ multiplied by $f_\phi$.\footnote{
Depending on the details of the inflaton potential, it can be a fraction of $\pi f_\phi$.
A more precise explanation will be given later.
}
We assume 
that the Hubble parameter $H_{\rm inf}$ during inflation is lower than the QCD scale so that the QCD axion acquires the potential during inflation. Then, if the inflaton has a mixing with the QCD axion\footnote{
The mass and kinetic mixings between the QCD axion and another axion were discussed in e.g. Refs.~\cite{Kitajima:2014xla, Daido:2015bva, Daido:2015cba,Higaki:2015jag, Ho:2018qur}, where the level crossing between the two axions can reduce the axion abundance. The mixing can also induce the DM decay~\cite{Higaki:2014qua,Daido:2017wwb,Kobayashi:2019eyg}. Also the mixing between axions plays an important role in inflation model building~\cite{Kim:2004rp, Choi:2014rja, Higaki:2014pja,Higaki:2014mwa,Choi:2015fiu,Kaplan:2015fuy}
}, the sign of the QCD axion potential is flipped by the phase shift of $\pi$ due to the inflaton dynamics. The role of the inflation model is twofold. First, the inflation scale is so low that the QCD axion 
 is naturally located near the potential minimum during inflation. The probability distribution of the QCD axion
 around the potential minimum is determined by the Bunch-Davies (BD) distribution~\cite{Graham:2018jyp,Guth:2018hsa}. Second, the inflaton dynamics provides the phase shift of $\pi$ in the axion potential through the mixing, and flips the sign of the potential. In other words, the initial position of the QCD axion is set close to the hilltop. We call such an inflation model that provides the phase shift of $\pi$ as $\pi$\hspace{-0.2mm}nflation. The $\pi$\hspace{-0.2mm}nflation resolves the two issues in realizing the QCD axion DM with small $f_a$ simultaneously.

In fact, the basic idea of the $\pi$\hspace{-0.2mm}nflation was proposed by Daido and the present authors in 2017~\cite{Daido:2017wwb} 
where it was shown that the sign of the potential can be flipped by the phase shift close to $\pi$ of the heavy axion via the 
mass mixing. The main purpose was to explain the initial condition for the hilltop inflation. 
Recently, we pointed out  in Ref.~\cite{Takahashi:2019qmh} that, if the inflaton has a mixing with the QCD axion, 
the minimum of the axion potential can be shifted after inflation, inducing coherent oscillations of the QCD axion at a later time 
when the Hubble parameter becomes comparable to the axion mass. In both cases, the axion mass was assumed to be much smaller
than the Hubble parameter during inflation. 
On the other hand, more recently, Kobayashi and Ubaldi showed in Ref.~\cite{Kobayashi:2019eyg}
that the axion that is already stabilized at the potential minimum during inflation can be produced by the inflaton dynamics.
The key differences are that they assumed that the axion is heavier than the Hubble parameter during inflation, 
and they made use of the kinetic mixing instead of the mass mixing.

Recently, Co, Gonzalez, and Harigaya proposed a mechanism that drives the QCD axion to the hilltop of the potential in a supersymmetric framework~\cite{Co:2018mho}. The QCD axion has a mass larger than $H_{\rm inf}$ in their set-up due to the enhanced QCD scale~\cite{Dvali:1995ce,Banks:1996ea,Choi:1996fs,Jeong:2013xta}, and it is stabilized at the potential minimum at that time.\footnote{
The axion isocurvature perturbations can be suppressed in this case~\cite{Jeong:2013xta}.
See e.g. Refs.~\cite{Linde:1990yj,Linde:1991km,Lyth:1992tw,Kasuya:1996ns,Dine:2004cq,
Folkerts:2013tua,Higaki:2014ooa,Dine:2014gba,Nakayama:2015pba,Harigaya:2015hha,
Choi:2015zra,Kawasaki:2015lea,Takahashi:2015waa,Agrawal:2017eqm,Kawasaki:2017xwt,Kitajima:2017peg,Tenkanen:2019xzn} for other scenarios
 to suppress the axion isocurvature. 
} 
Their mechanism flips the coefficient of the axion potential. On the other hand,
in our scenario, the potential is flipped by the phase shift of $\pi$ by our $\pi$\hspace{-0.2mm}nflation. 

We note that, if  the QCD axion is exactly on top of the potential when it starts to oscillate during the QCD phase transition, domain walls will be produced.\footnote{
FT thanks Keisuke Harigaya for discussion on this point.}
Such domain walls without strings are stable and spoil the success of the standard cosmology. The cosmological catastrophe may be avoided if the initial position is slightly deviated from the CP conserving minimum (or maximum). In our scenario,  the QCD axion remains light and follows the BD distribution during inflation. Thus, it is naturally deviated from the potential minimum by a small amount which is determined by the inflation scale.

The rest of this paper is organized as follows. In the Sec.\,\ref{sec:2} we present the basic idea of $\pi$\hspace{-0.2mm}nflation and 
show that the QCD axion with $f_a \ll 10^{12}$\,GeV can explain DM due to $\pi$\hspace{-0.2mm}nflation.  In Sec.\,\ref{sec:3} we provide a 
concrete $\pi$\hspace{-0.2mm}nflation model and study its experimental and observational 
implications. The last section is devoted to discussion and conclusions.

\section{Basic idea}
\label{sec:2}
In this section we explain the basic idea of the $\pi$\hspace{-0.2mm}nflation and its implications for the QCD axion abundance
and isocurvature perturbations. 

\subsection{$\pi$\hspace{-0.2mm}nflation}
Let us introduce an inflaton field, $\phi$, and the QCD axion, $a$.\footnote{
Our mechanism works even if $\phi$ is not the inflaton,
as long as $\phi/f_\phi$ changes its field value by $\pi$ mod $2\pi$ after inflation.
} Both fields enjoy the following discrete shift symmetry,
\begin{align}
\laq{shift}
\f &\rightarrow \f+2\pi f_\f,~~~
a\rightarrow a+2\pi f_a,
\end{align}
where $f_\phi$ and $f_a$ are the decay constants of $\phi$ and $a$, respectively. 
This implies that the potential is periodic with respect to both $\phi$ and $a$. 

In order not to spoil the Peccei-Quinn mechanism
as a solution to the strong CP problem,
there must be a flat direction when one turns off the QCD interactions. Then, after a certain field redefinition,
the potential can be given by the following form,
\beq \laq{tot}
V(\f, a)= V_{\rm inf}\(\phi\)+ \chi(T)\left[1-\cos{\(n_{\rm mix} \frac{ \f - \phi_{\rm min}}{f_\f}+\frac{a}{f_a}\)}\right],
\eeq
where the first term is the inflaton potential $V_{\rm inf}$ satisfying $V_{\rm inf}(\phi + 2\pi f_\phi) = V_{\rm inf}(\phi)$,
 $\chi(T)$ is the topological susceptibility of QCD, and an integer, $n_{\rm mix}$, represents the anomaly coefficient 
 of the mixing term. 
 The mixing between $\phi$ and $a$ is obtained if both $\phi$ and $a$ couple to the gluon field strength, $G,$ and its dual, $\tl{G}$, as 
\beq
\laq{Lag}
{\cal L}\supset \frac{\a_s}{8\pi}\(n_{\rm mix}\frac{\f}{f_\f}+\frac{a}{f_a}\)G\tl{G},
\eeq
where $\a_s$ is the strong coupling constant.

 We assume that $V_{\rm inf}(\phi)$ has a flat plateau in some finite neighborhood of $\phi = \phi_{\rm inf}$ where
 eternal inflation~\cite{Linde:1982ur,Steinhardt:1982kg,Vilenkin:1983xq} takes place. The size of such region is
 assumed to be so small that it has no effect on the determination of the
 QCD axion abundance. This is indeed the case in the inflation model to be discussed 
 in the next section.
 After inflation ends, the inflaton is stabilized at the nearest minimum, $\phi = \phi_{\rm min}$, where $V_{\rm inf}(\phi_{\rm min}) \simeq 0$.
 
In the present vacuum, we assume that the inflaton is much heavier than the QCD axion so that the inflaton 
can be safely integrated out. Then, the topological susceptibility is related to the QCD axion mass 
in a usual way as
$\chi(T) = m_a(T)^2 f_a^2$, where $m_a(T)$ is the temperature-dependent axion mass given by
\begin{equation}
\label{mass}
m_a(T) \;\simeq\;
\begin{cases}
\displaystyle{\frac{\sqrt{\chi_0}}{ f_a}} \left(\frac{T_{\rm QCD}}{T}\right)^n &~~T \gtrsim T_{\rm QCD}\vspace{3mm}\\
\displaystyle{5.7 \times 10^{-6} \(\frac{10^{12}\GEV}{f_a}\)  {\rm eV}}&~~ T\lesssim  T_{\rm QCD}
\end{cases},
\end{equation}
with $n \simeq 4.08$~\cite{Borsanyi:2016ksw},  $T_{\rm QCD}\simeq 153 \MEV$ and $\chi_0 \simeq \(75.6 \MEV\)^4$. 

During inflation, on the other hand, there are two things to watch out for. One is that 
the QCD axion mass during inflation depends on the Gibbons-Hawking 
temperature~\cite{Gibbons:1977mu}
 \beq
 \label{gh}
  T_{\rm inf}\equiv \frac{H_{\rm inf}}{2\pi},
  \eeq 
  where 
 $H_{\rm inf}\simeq\sqrt{V_{\rm inf}(\f_{\rm inf})/3M_{\rm pl}^2}$ is the Hubble parameter during inflation,
 and $M_{\rm pl} \simeq 2.4\times 10^{18}\GEV$ is the reduced Planck mass. 
 Therefore, for $T_{\rm inf} \lesssim T_{\rm QCD}$, or equivalently, $H_{\rm inf} \lesssim 1\,$GeV, 
 the QCD axion acquires its potential during inflation.\footnote{
 For $T_{\rm inf} \gtrsim T_{\rm QCD}$, the use of Eq.~(\ref{mass}) during inflation should be taken with a grain of salt,
 because the Hawking radiation in the de Sitter space is not exactly same as in the lattice calculation.
 }
 The other thing is that the QCD axion $a$ and the inflaton $\phi$ are not the mass eigenstates, in general.
 In this case one has to take account of the mixing between them to follow their evolution during inflation. 
For simplicity we assume the absolute magnitude of the curvature along the $\phi$ direction is so large that
the mixing is negligibly small, i.e.,
\beq 
\laq{cond1}
\sqrt{|V_{\rm inf}''(\phi_{\rm inf})|}
\gg n_{\rm mix}\frac{\sqrt{\chi(T_{\rm inf})}}{f_\f}.
\eeq
Since $H_{\rm inf}^2 \gtrsim |V_{\rm inf}''(\phi_{\rm inf})|$ is required for the slow-roll inflation,
the above condition implies $V_{\rm inf}(\phi_{\rm inf})\gg \chi(T_{\rm inf})$ for $f_\f < M_{\rm pl}$ and $n_{\rm mix}
= {\cal O}(1)$. In other words, the inflation is mainly driven by  $V_{\rm inf}$ not the potential induced by non-perturbative QCD effect. 
Then,  the QCD axion $a$ is almost the lighter mass eigenstate whose mass satisfies
\begin{align}
\label{maH}
m_a(T_{\rm inf}) \ll H_{\rm inf}.
\end{align}
This should be contrasted to 
Refs.~\cite{Dvali:1995ce,Banks:1996ea,Choi:1996fs,Jeong:2013xta,Co:2018mho}.\footnote{
\label{f6}
If $m_a(T_{\rm inf}) \gtrsim H_{\rm inf}$, the QCD axion will generically follow the shift of the potential minimum,
and the hilltop initial condition is not realized. One needs some contrivance to realize the hilltop initial condition in this case;
e.g. the potential changes much faster than $m_a(T_{\rm inf})^{-1}$ after inflation,
or the QCD axion potential vanishes due to the temporal increase of the temperature 
(e.g. by decays of heavy particles) around the end of inflation.
 }

The mixing between $\phi$ and $a$ is small both during inflation and in the low energy, but the inflaton dynamics does
contribute to the effective strong CP phase and shifts the potential minimum of $a$~\cite{Daido:2017wwb,Takahashi:2019qmh}. In general, we define
 {\it $\pi$\hspace{-0.2mm}nflation} such that the field evolution of the inflaton (i.e. $\pi$\hspace{-0.2mm}nflaton)
gives a phase shift close to $\pi$ (mod $2\pi$) for another axion through mixing. 
In our case, this is equivalent to
\beq
\laq{cond}
n_{\rm mix}\frac{(\f_{\rm min}-\f_{\rm inf})}{f_\f}= {\pi+\delta} \mod {2\pi},
\eeq
with
\beq
|\delta|\ll 1, \non
\eeq
where the precise value of $\d$ is determined once the $\pi$\hspace{-0.2mm}nflation model is given. 
 In fact, we have $\f_{\rm min}-\f_{\rm inf} = \pi f_\phi$ in the $\pi$\hspace{-0.2mm}nflation model 
to be given in the next section. 
Thus, the minimum of the QCD axion potential  is located at
\begin{align}
\frac{a_{\rm min}^{\rm (inf)}}{f_a} = \pi - \delta~\mod {2\pi}
\end{align}
during inflation, while the minimum is shifted to
\begin{align}
\frac{a_{\rm min}^{\rm (vac)}}{f_a} = 0
\end{align}
after inflation. Here
we consider the range of $-\pi < a/f_a \leq \pi$ without loss of generality.
In other words, the potential minimum during inflation 
turns into (almost) the maximum after inflation. This is what the $\pi$\hspace{-0.2mm}nflation does. Now the question is the dynamics of
 the QCD axion, which will be studied in the next subsection.

Lastly let us comment on a possible contribution of the stochastic dynamics of  $\phi$ to the
phase shift. We assume that the eternal inflation takes place in the finite neighborhood of $\phi_{\rm inf}$,
$\phi = \phi_{\rm inf} \pm \Delta \phi_{\rm st}$.  In principle, the probability distribution of the inflaton in this region 
could contribute to $\delta$. However, in the $\pi$\hspace{-0.2mm}nflation model we consider,
$\Delta \phi_{\rm st}$ is not many orders of magnitude larger than $H_{\rm inf}$, and so, 
it is smaller than the other contributions. Moreover, the precise probability distribution of $\phi$ within $|\phi - \phi_{\rm inf}| < \Delta \phi_{\rm st}$
depends on the volume measure. Therefore, we do not consider the contribution of the stochastic dynamics
of $\phi$ to $\delta$ in the following.

\subsection{Initial misalignment angle}

During inflation the axion $a$ remains light and acquires quantum fluctuations.
If the inflation lasts long enough, more specifically, if the number of $e$-folds satisfies  
$N\gg H_{\rm inf}^2/m_a^2$, the quantum diffusion is balanced by the classical motion,
and the axion field distribution asymptotes to the BD distribution peaked at the 
potential minimum during inflation, $a=a_{\rm min}^{\rm (inf)}$,
with the variance~\cite{Graham:2018jyp,Guth:2018hsa}
\beq
\laq{BDdis}
\vev{\left(a-a^{\rm (inf)}_{\rm min}\right)^2} \;\simeq\; \frac{3H_\text{inf}^4}{8\pi^2 m^2_a(T_{\rm inf}) }.
\eeq
Here the average is taken over superhorizon patches, and 
we approximated the axion potential by a quadratic term assuming $H_{\rm inf}^2 \lesssim m_a(T_{\rm inf}) f_a$.
Thus, the typical initial misalignment angle set during inflation is given by
\beq
\laq{dtheta}
\ab{\theta_i-\pi}\simeq \max\left[\sqrt{ \frac{3H_\text{inf}^4}{8\pi^2\chi(T_{\rm inf})}},|\d|\right]\ll 1,
\eeq
barring cancellation between the two contributions. We emphasize here that, even for $\delta = 0$, the axion is not
exactly at the minimum $a^{\rm (inf)}_{\rm min}$, thereby avoiding the aforementioned domain-wall problem. 

Soon after the inflation ends, $\f$ starts to oscillate around the potential minimum~$\f=\f_{\rm min}$.
and decays into the standard model (SM) particles to reheat the universe. Note that
the inflaton is coupled to gluons via the mixing with $a$, and it may have other interactions as well.
Depending on the couplings, the inflaton can decay and evaporate very efficiently. 
The inflaton decay soon produces radiation with temperature greater than 
 $T_{\rm QCD}$, and hence the QCD axion potential vanishes.  
 Since the axion mass is much smaller than the Hubble parameter as in
 Eq.~(\ref{maH}),
the axion field hardly moves from the initial position $a_i = \theta_i f_a$ until it starts to oscillate 
when  the QCD axion potential is generated again during the QCD phase transition. 
Therefore, the initial misalignment angle of the QCD axion is given by \Eq{dtheta}, and it is close to the
hilltop thanks to the $\pi$\hspace{-0.2mm}nflation.\footnote{We assume here that the phase shift takes place instantaneously after inflation. In realistic inflation models, however,
the axion field value may receive a small correction of order $\delta \theta_i \sim m_a^2/H_{\rm inf}^2$. This only slightly increases
the lower bound on $f_a$  from $3 \times 10^9$\,GeV to $(4-5) \times 10^9$\,GeV.
}

\subsection{Axion abundance and isocurvature bound}

In the $\pi$\hspace{-0.2mm}nflation scenario, the initial position of the QCD axion is naturally set around the potential maximum, 
which delays the onset of oscillations, thereby increasing the abundance. The $\pi$\hspace{-0.2mm}nflation enables
 the QCD axion  to explain DM  even for  $f_a \ll 10^{12}\GEV$.

\begin{figure}[!t]
\begin{center}  
   \includegraphics
[width=105mm]{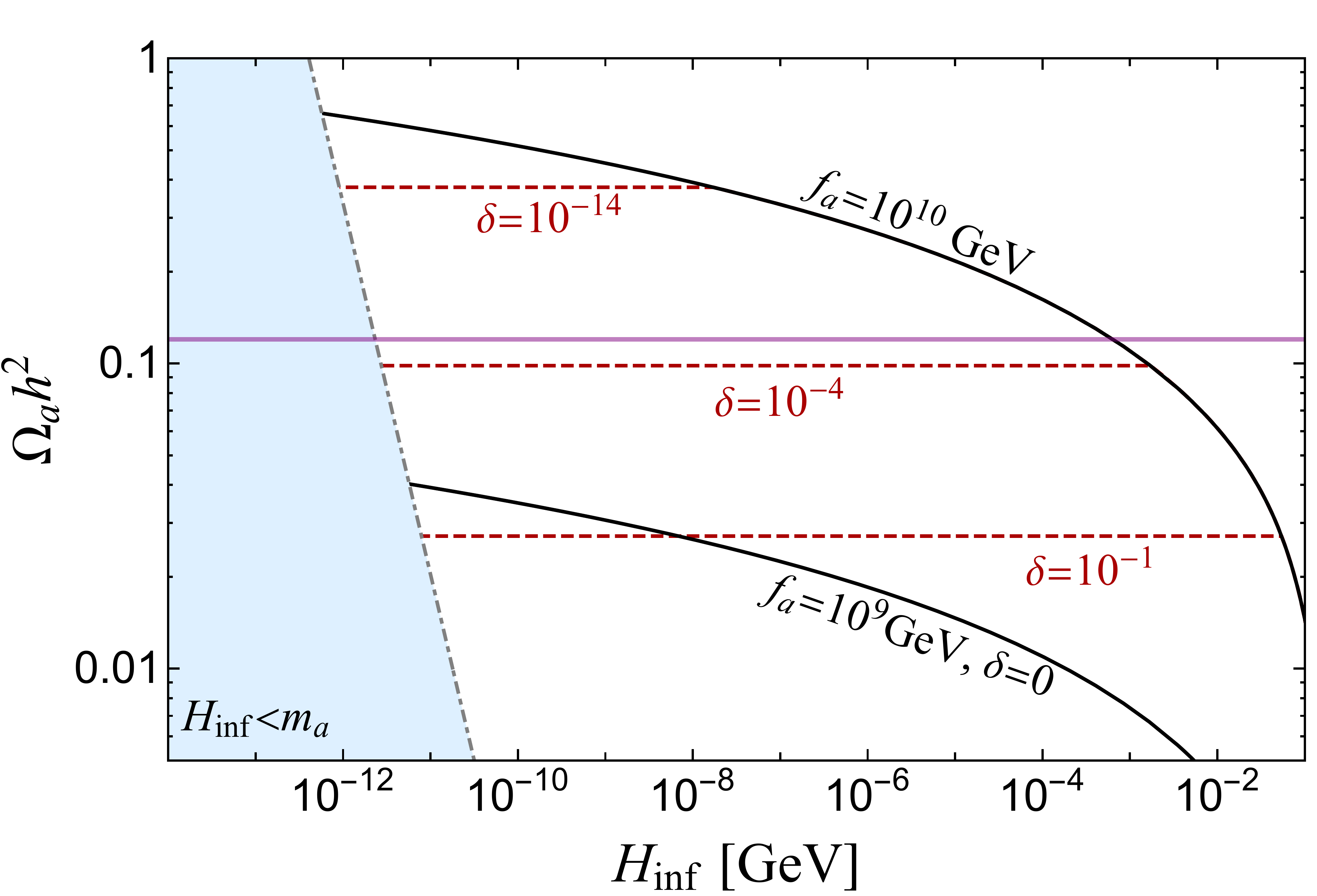}
      \end{center}
\caption{
$\Omega_a h^2$ with  the initial misalignment angle 
\eq{dtheta} and $\delta = 0$, as a function of $H_{\rm inf}$ for $f_a = 10^9$ and $10^{10}$\,GeV. 
We also show the cases of $\delta = 10^{-1}, 10^{-4}, 10^{-14}$ for $f_a = 10^{10}$\,GeV in red  dashed  lines, from bottom to top. 
In the left blue shaded region, the axion is not light, i.e., $m_a(T_{\rm inf}) > H_{\rm inf}$,
 and our analysis cannot be applied.  The  narrow horizontal purple band represents the observed DM density.  }\label{fig:1} 
\end{figure}

In Fig.\,\ref{fig:1} we show the abundance of the QCD axion for the initial misalignment angle 
\eq{dtheta} with $\delta = 0$, as a function of $H_{\rm inf}$  for $f_a = 10^{9}$ and $10^{10}$\,GeV. We also show
the cases of $\delta = 10^{-1}, 10^{-4}, 10^{-14}$ for $f_a = 10^{10}$\,GeV as the horizontal lines from bottom to top.  The behavior of these lines can be understood as follows. 
First let us suppose $\delta = 0$.
The typical deviation from the potential maximum is given by \Eq{dtheta}, and it 
decreases as $H_{\rm inf}$ decreases. As a result, the axion abundance increases for a fixed $f_a$.
This behavior is represented by the black solid lines  for $f_a = 10^{9}$ and $10^{10}$\,GeV.
On the other hand, if $\delta \ne 0$, the deviation from the potential maximum will be determined by $\delta$ for a sufficiently
small $H_{\rm inf}$, and then, the axion abundance becomes independent of $H_{\rm inf}$, 
which is represented by the horizontal lines for different values of $\delta$. 
The narrow horizontal (purple) line represents the observed DM abundance. 
In the left shaded (blue) region, the axion is heavy during inflation, i.e.,  $m_a(T_{\rm inf}) > H_{\rm inf}$, and
our approximation breaks down. One needs to follow the evolution of the QCD axion in this case (cf. the footnote \ref{f6}).
One can see from the figure that the QCD axion can explain DM, for instance, with $f_a = 10^{10}$\,GeV for $H_{\rm inf} \simeq 1$\,MeV
or for $H_{\rm inf} \lesssim 1$\,MeV and $\delta \simeq 10^{-4}$.

The axion isocurvature perturbation is also enhanced in the hilltop limit~\cite{Lyth:1991ub,Kobayashi:2013nva}.\footnote{
The non-Gaussianity is mildly enhanced~\cite{Kobayashi:2013nva}, but it is estimated to be well within
the current bound~\cite{Akrami:2019izv}. 
}
The isocurvature perturbation is given by~\cite{Kobayashi:2013nva}
\beq
S = \frac{\Omega_a}{\Omega_{\rm DM}}  \frac{H_{\rm inf}}{2\pi f_a} \frac{\partial}{\partial \theta_i} \log{[\Omega_a(\theta_i) ]}.
\eeq
Using \eq{ab}, one can see that the isocurvature perturbation gets significantly enhanced as
\beq
\label{Siso}
S \propto \frac{ H_{\rm inf}}{ |\theta_i-\pi|},
\eeq
in the hilltop limit $|\theta_i-\pi| \ll 1$, where we have dropped the logarithmic dependence.
Its power spectrum ${\cal P}_S$ should satisfy the CMB bound~\cite{Akrami:2018odb} 
\beq
{\cal P}_S \; \lesssim\; 8.4\times 10^{-11},
\eeq
and only low-scale inflation models are allowed. This is consistent with our assumption
that the QCD axion acquires the potential during inflation (see discussion below (\ref{gh})).

\begin{figure}[!t]
\begin{center}  
     \includegraphics
[width=105mm]{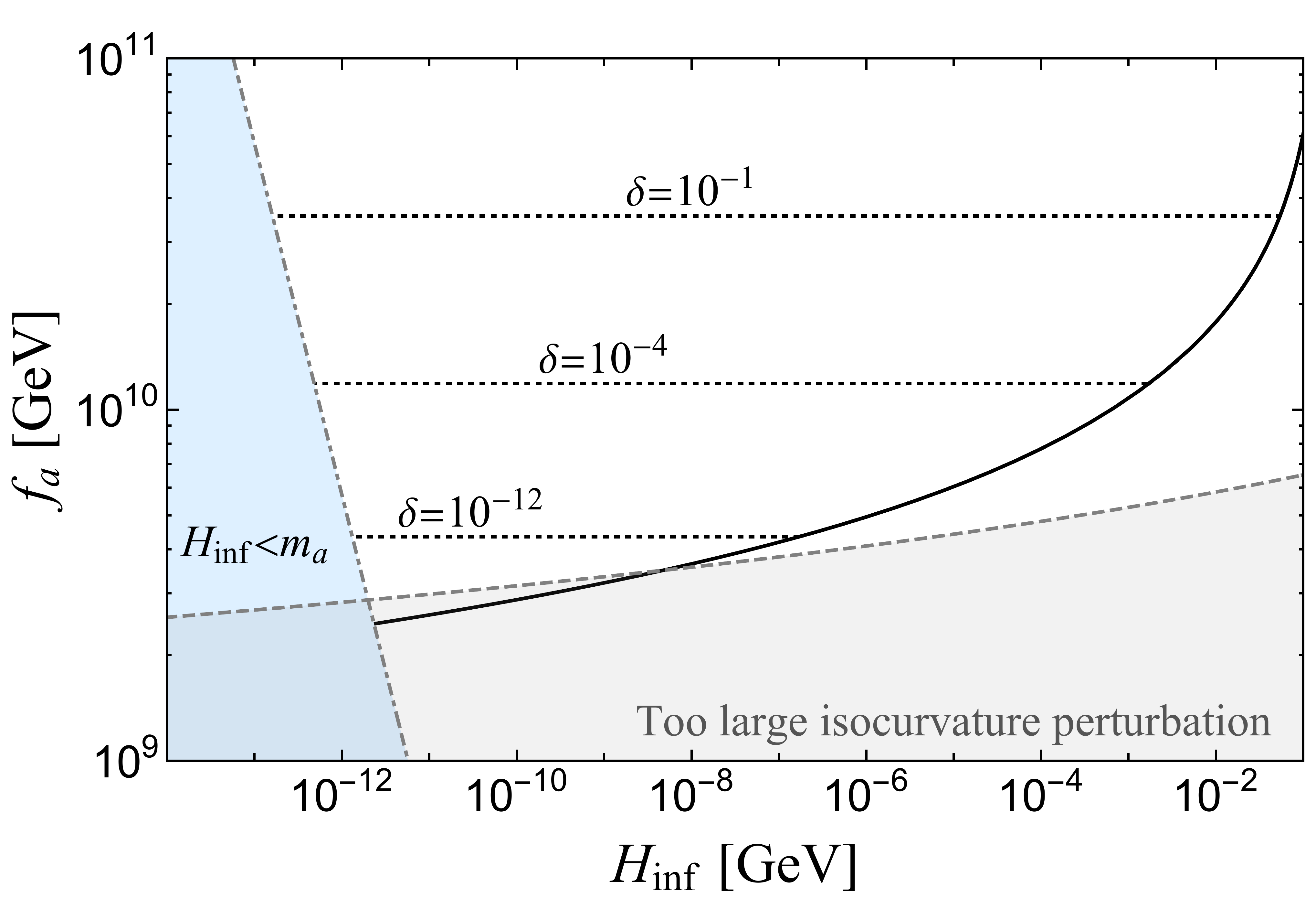}
      \end{center}
\caption{$f_a$ as a function of $H_{\rm inf}$ for $\delta = 0, 10^{-12}, 10^{-4}, 10^{-1}$, where the
QCD axion with the initial misalignment \eq{dtheta} explains DM.
The lower  gray region is excluded due to too large isocurvature perturbation. 
In the left blue region, $m_a(T_{\rm inf})>H_{\rm inf}$, and our analysis cannot be applied.}\label{fig:2} 
\end{figure}

In Fig.~\ref{fig:2}, we show the value of $f_a$ for the axion to explain DM as a function of $H_{\rm inf}$
in the case of $\delta =  0, 10^{-12}, 10^{-4}, 10^{-1}$, where the initial misalignment angle is given by
\Eq{dtheta}. For $\d=0$, \Eq{dtheta} gives $|\theta_i -\pi| \propto H_{\rm inf}^2$, 
thus ${\cal P}_S\propto H_{\rm inf}^{-2}$ for the fixed axion abundance. In other words,
the isocurvature perturbation gets enhanced for smaller values of $H_{\rm inf}$,
which can be seen in  Fig.~\ref{fig:2} by noting that a part of the  black solid line is within
the gray shaded region for $H_{\rm inf} \lesssim 10^{-8}$\,GeV.
This should be contrasted to the usual case where the isocurvature can be avoided
for sufficiently small $H_{\rm inf}$.
Therefore, $H_{\rm inf} \lesssim 10^{-8}$\,GeV, one needs a nonzero $\d$ to satisfy the isocurvature bound.
The precise value of $\d$ depends on details of the $\pi$\hspace{-0.2mm}nflation.
In the next section we will see that such nonzero $\d$ is required to explain the observed spectral index
in the axion-like-particle (ALP) $\pi$\hspace{-0.2mm}nflation.
The decay constant is also bounded from below,
\beq
f_a \gtrsim 2.9 \times 10^9\GEV 
\eeq 
or equivalently, the QCD axion mass is bounded above,
\beq
m_{a} \lesssim 2.0\times 10^{-3}\EV.
\eeq
This is the robust prediction of the hilltop QCD axion DM with the $\pi$\hspace{-0.2mm}nflation.\footnote{
In the left blue region, the axion field generically evolves after inflation, and the hilltop initial condition is
no longer realized. Thus, even larger $f_a$ will be required to explain DM.
}

The QCD axion DM of mass in the range of $10^{-4} \text{--} 10^{-3}$\,eV can be searched for in the axion haloscopes
such as MADMAX~\cite{TheMADMAXWorkingGroup:2016hpc, Brun:2019lyf}
and TOORAD~\cite{Marsh:2018dlj} experiments, where the axion-photon coupling is assumed to be the one predicted by the
KSVZ QCD axion model~\cite{Kim:1979if,Shifman:1979if}.
If the coupling of the axion to photons is slightly enhanced by $\O(1)$ factor,\footnote{This is the case if the PQ quarks have slightly large electric charge. Alternatively, the photon coupling can be enhanced due to the mixing between the photon and hidden photon which leads to gauge coupling unification~\cite{Daido:2018dmu}. The coupling of the axion to gauge bosons can be highly enhanced by the clockwork mechanism~\cite{Higaki:2016yqk}, and so, the axion-photon coupling can be similarly enhanced~\cite{Farina:2016tgd}. } such axion can also be searched for in the ORGAN~\cite{McAllister:2017lkb} and IAXO experiments~\cite{Irastorza:2011gs, Armengaud:2014gea, Armengaud:2019uso}.

\section{Concrete $\pi$\hspace{-0.2mm}nflation models} 
\label{sec:3}
In this section we first provide successful  $\pi$\hspace{-0.2mm}nflation
in which the inflaton potential consists of multiple cosine functions satisfying the
discrete shift symmetry \eq{shift}. Then we estimate the $\delta$ in this model,
which determines the initial misalignment angle of the QCD axion. 
Finally we discuss the experimental and observational implications.

\subsection{Model}
The inflaton potential $V_{\rm inf}(\phi)$ remains unchanged under the shift of
the inflaton, \eq{shift}. Such periodic potential can be expanded as a
Fourier series. If a single cosine function gives the dominant contribution,
it is the so-called natural inflation~\cite{Freese:1990rb,Adams:1992bn},
which necessitates a decay constant $f_\phi$
of order or larger than the Planck scale for slow-roll. 
However, the natural inflation is already disfavored by the current 
CMB observation~\cite{Akrami:2018odb}. Moreover, the predicted inflation scale
is much higher than the QCD scale, and so, it does not fit our purpose.\footnote{ 
If we abandon to explain the observed density perturbation with the single inflation,
the natural inflation with $H_{\rm inf} \lesssim 1$\,GeV
can be the $\pi$\hspace{-0.2mm}nflation. In this case, we need another short inflation 
(or curvaton) which generates the primordial density perturbation with the right magnitude. 
}

Let us here consider the so-called multi-natural inflation~\cite{Czerny:2014wza, Czerny:2014xja,Czerny:2014qqa,Higaki:2014sja}, 
where multiple cosine terms conspire to realize a sufficiently flat potential. The multi-natural inflation
works even for sub-Planckian decay constants, and we assume $f_\phi \ll M_{\rm pl}$ in the following. 
The flat-top potential with multiple cosine terms  has 
several possible UV origins e.g. in 
supergravity\cite{Czerny:2014xja,Czerny:2014qqa,Higaki:2014sja} and
extra dimensions~\cite{Croon:2014dma}. A similar potential with an elliptic function
is also obtained at the low-energy limit of string-inspired setups~\cite{Higaki:2015kta, Higaki:2016ydn}. 

We focus on the minimal case in which the potential is dominated by the two cosine terms,
\begin{align}
\label{eq:DIV} 
V_{\rm inf}(\phi) = \Lambda^4\(\cos\(\frac{\phi}{f_\f} + \Theta \)- \frac{\kappa }{n_{\rm inf}^2}\cos\(n_{\rm inf}\frac{\f}{f_\f }\)\)+{\rm 
const.}.
\end{align}
where $n_{\rm inf}$ ($>1$) is an integer,  $\kappa$ and $\Theta$ parameterize the relative height and phase of the two terms, 
respectively, and the last constant term is introduced to make the cosmological constant vanishingly small in the 
present vacuum.\footnote{
One can consider a case in which the first cosine term in \Eq{DIV} contains
another positive integer $n'_{\rm inf} < n_{\rm inf}$. It is straightforward to extend our analysis to this case by redefining the decay constant. 
In particular, $\f_{\rm inf}/f_\f-\f_{\rm min}/f_\f\simeq \pi /n'_{\rm inf} \mod {2\pi}$ can be simply obtained if $n_{\rm inf}/n'_{\rm inf}$ is an integer. Thus, the $\pi$\hspace{-0.2mm}nflation defiend in \eq{cond} can have even $n_{\rm mix}$.} 
A relative CP phase $\Theta$ can be naturally nonzero if the two terms originate from different sources, and its typical value
depends on the UV completion.  In fact, as we shall see shortly, $\Theta$ can be fixed by the observed
spectral index.

In the limit of $\Theta=0$ and $\kappa = 1$, the above potential is reduced to the hilltop quartic inflation 
model where the inflation takes place in the neighborhood of $\f_{\rm inf}= 0$.\footnote{
From the low-energy point of view, we do not find any particular reason to set $|\Theta| \ll 1$ and $\kappa \approx 1$
other than the requirement for successful slow-roll inflation.
} 
Without loss of generality, we assume that the inflaton field value increases during the slow-roll. Then, after
inflation, the inflaton will be stabilized at the nearest potential minimum,  $\f_{\rm min}=\pi f_a$.
Therefore, this inflation model satisfies \Eq{cond} as long as $\Theta \approx 0$ and $\kappa \approx 1$.
The multi-natural inflation can easily be $\pi$\hspace{-0.2mm}nflation if $n_{\rm mix}$ is odd. 
For simplicity, in the following discussion we take 
\beq
n_{\rm mix}=1.
\eeq
As we will see, 
a nonzero $\Theta$ is required to explain the observed spectral index, which is essential for determining 
$\d$.

\subsection{Inflaton dynamics}
Here we briefly review the inflation dynamics with the potential \eq{DIV}. For more 
detailed analysis, see Refs.\,\cite{Daido:2017wwb, Daido:2017tbr, Takahashi:2019qmh}.
Let us expand the potential around $\f=0$, 
\begin{equation}
\label{eq:app}
V_{\rm inf}(\phi) \simeq V_0   - \Theta \frac{\L^4 }{ f_\f} {\f} +\frac{m^2 }{2}\f^2 - \lambda \f^4+\cdots,    
\end{equation}
where $\cdots$ represents terms with negligible effects on the inflaton dynamics during  inflation.
Here we have defined 
 \begin{align}
 \label{eq:V0}
 V_0 & \equiv \left(2 - \frac{2}{n_{\rm inf}^2} \sin^2{\frac{n_{\rm inf} \pi}{2}} \right) \Lambda^4,\\
 m^2 &\equiv {\(\kappa-1\)}\frac{\L^4 }{f_\f^2},\\
 \lambda &\equiv \frac{n_{\rm inf}^2-1 }{4!}\(\frac{\Lambda}{f_\f}\)^4.
 \label{eq:lambda}
\end{align}
Note that $V_0$ is chosen so that the potential vanishes at the minimum,
$\phi_{\rm min} \simeq \pi f_\f$.
Obviously, the potential \eq{app} is reduced to that of the hilltop quartic inflation
in the limit of $\Theta \rightarrow 0$ and $\kappa \rightarrow 1$. 
In this limit, the potential is extremely flat around the origin where the eternal inflation
takes place. This remains to be the case if $\kappa$ and $\Theta$ are in the following
range\cite{Takahashi:2019qmh}\footnote{$\H$ generally exists in the potential. The tiny value may be due to the non-perturbative dynamics (cf. CKM phase contribution to the QCD axion potential). It may also be the expectation value of another axion. In particular, it may be another light axion following the BD distribution. }
\beq
\label{cfei}
 |\k-1|\lesssim \(\frac{f_\f}{M_{\rm pl}}\)^2,|\Theta|\lesssim \(\frac{f_\f}{M_{\rm pl}}\)^3.
\eeq
The Hubble parameter during the eternal inflation is given by 
\beq \laq{hinfV0} H_{\rm inf}\simeq \sqrt{\frac{V_0}{3 M_{\rm pl}^2 }}.\eeq 
Note that the eternal inflation may explain the initial condition for the hilltop inflation, and moreover,
it plays an essential role to realize the BD distribution for the QCD axion.

The present $\pi$\hspace{-0.2mm}nflation model is well approximated by a simple hilltop quartic inflation,
and so, let us set $\Theta = 0$ and $\kappa = 1$ for the moment. The effects of $\Theta \ne 0$ and 
$\kappa \ne 1$ are taken into account in our numerical calculations.
The quartic coupling is fixed by the CMB normalization of the primordial density perturbation,
\beq
\laq{quo}
\lambda\simeq 2.9\times 10^{-13} \(\frac{30}{N_*}\)^{3},
\eeq
where $N_*$ is the e-folding number at the horizon exit of the CMB scales, given by
\begin{equation}
\label{eq:efold}
N_*\simeq 28+\log{\(\frac{V_0^{1/4} }{10\,\TEV}\)}.
\end{equation}
Here we have assumed the instantaneous reheating. 
As we are interested in $H_{\rm inf}\lesssim 1\GEV$, \Eq{efold} implies $N_{*}\lesssim 40$.
The inflaton $\f$ has a coupling to gluons
through the mixing with the QCD axion, and it may also have other interactions with the SM particles. 
The reheating proceeds through both perturbative decay and
dissipation effects. The reheating is indeed instantaneous in a wide parameter range,
especially if the inflaton is coupled to the top quarks~\cite{Daido:2017tbr,Takahashi:2019qmh}.

 From the CMB normalization \eq{quo}, one obtains 
\beq
\label{eq:ratio}
\frac{\Lambda}{f_\f}\simeq  1.2\times 10^{-3 }\(\frac{3}{n_{\rm inf}^2-1}\)^{\frac{1}{4}}\(\frac{30}{ N_*}\)^{\frac{3}{4}}.
\eeq
In the case of even $n_{\rm inf}$, this relates the mass of the inflaton at the potential minimum to 
the decay constant $f_\f$, and it is given by~\cite{Czerny:2014xja,Takahashi:2019qmh}
\begin{align}
\label{mphi}
m_\phi \simeq  \sqrt{2} {\frac{\Lambda^2}{f_\f}} \sim 10^{-6} f_\phi~~~[{\rm for~even}~n_{\rm inf}],
\end{align}
where we have used $\Theta\simeq 0$ and $\kappa \simeq 1$.
In the case of odd $n_{\rm inf}$, the inflaton mass becomes much smaller
due to the upside-down symmetry. We will return to this case later.

The scalar spectral index $n_s$ in the hilltop quartic inflation is predicted to be
\beq
n_s\simeq 1-{\frac{3}{N_*}},
\eeq
which is too small to explain
the observed scalar spectral index, $n_s^{\rm CMB} = 0.9649\pm0.0042$ \cite{Akrami:2018odb},
for  $N_* \lesssim 40$.
In fact, it is known that $n_s$ is rather sensitive to possible small corrections to the inflaton potential, 
and one can easily increase the predicted value of $n_s$ to give a better fit to the CMB data by introducing 
small but non-zero $\Theta$~\cite{Takahashi:2013cxa}.
Introducing a nonzero $\kappa-1$ has a similar but slightly weaker effect.
One can explain the observed $n_s$ by introducing a non-vanishing $\H >\O(0.001) \(f_\f/M_{\rm pl}\)^3$~\cite{Daido:2017wwb,Daido:2017tbr}.
By combining this with the condition (\ref{cfei}), one arrives at 
\beq \Theta=\xi\frac{f^3_\f}{M_{\rm pl}^3},\eeq
with
\beq
\label{range}
\xi=\O(0.001-1).
\eeq
The nonzero $\Theta$ will be important for determining $\d$ in this $\pi$\hspace{-0.2mm}nflation.

\subsection{Prediction for initial misalignment angle }
The field excursion of the inflaton, $\phi_{\rm min} - \phi_{\rm inf}$, 
determines the phase shift of the QCD axion potential. In the model of (\ref{eq:DIV}) or (\ref{eq:app}),
the eternal inflation takes place in the vicinity of $\phi = \phi_{\rm inf}$ given by
\beq 
\frac{\f_{\rm inf}}{f_\f} \;\simeq\; -\(\frac{6\x}{n_{\rm inf}^2-1}\)^{1/3} \frac{f_\f}{M_{\rm pl}}
\eeq 
for $\kappa = 1$. Even if we vary $\kappa$ in the range of (\ref{cfei}), the result  only changes
by a factor of order unity.
On the other hand, 
the field value at the potential minimum is given by
\begin{align}
\frac{\f_{\rm min}}{f_\phi}=\pi  + \O\left(\left(\frac{f_\f}{M_{\rm pl}}\right)^3\right)
\end{align}
for even $n_{\rm inf}$.
Therefore, one obtains
\beq
\label{delta}
\d\simeq \(\frac{6\x}{n_{\rm inf}^2-1}\)^{1/3} \frac{f_\f}{M_{\rm pl}} ~~~[ {\rm for ~even~}n_{\rm inf}].
\eeq
Using Eqs.~\eq{dtheta}, (\ref{eq:V0}), \eq{hinfV0}, (\ref{eq:ratio}),  and (\ref{delta}), one arrives at
\begin{align}
\label{result}
\ab{\theta_i-\pi}\simeq 
2.4\times 10^{-9}   \left(\frac{N_*}{30}\right) \(\frac{\x^4}{n_{\rm inf}^2-1} \)^{1/9} 
 \(\frac{H_{\rm inf}}{H_{\rm inf}^{\rm T}}\)^p ~~~[ {\rm for ~even~}n_{\rm inf}].
\end{align}
with
\begin{align}
H_{\rm inf}^{\rm T} \;\simeq\; 8.3 \times 10^{-6}\GEV \left(\frac{N_*}{30}\right)^{1/2}   \(\frac{\x^4}{n_{\rm inf}^2-1} \)^{1/18}.
\end{align}
Here $p = 2$ for $H_{\rm inf}\gtrsim H_{\rm inf}^{\rm T}$ and $p=1/2$ for $H_{\rm inf}\lesssim H_{\rm inf}^{\rm T}$.
Therefore, the hilltop initial condition for the QCD axion can indeed be realized by the $\pi$\hspace{-0.2mm}nflation
model. 
It is interesting to note that, for $H_{\rm inf}<H_{\rm inf}^{\rm T}$, the axion DM abundance is determined by $\delta$, which is correlated with the observed spectral index. The isocurvature perturbation
behaves as
\beq
\label{isoeven}
{\cal P}_{S}\propto H_{\rm inf}^{2-2p},
\eeq
which takes the maximum value at $H_{\rm inf} \simeq H_{\rm inf}^{\rm T}$, but 
it is well below the current upper bound, as we shall see shortly.

In the case of  odd $n_{\rm inf}$, the situation is quite different. 
This is because the potential is upside-down symmetric,
\beq
\label{eq:udsym}V_{\rm inf}(\phi)=-V_{\rm inf}(\phi+\pi f_\f )+{\rm const.}~~~ [{\rm for~odd~} n_{\rm inf}],
\eeq
which implies that the potential minimum exactly differs from the maximum by $\pi f_\f.$
As a result, one finds
 \beq 
 \label{d0}
\d=0 ~~[{\rm for~odd}~n_{\rm inf}]
\eeq 
for any $\Theta$ and $\k-1$ in the range of (\ref{cfei}).
Thus, the deviation from the potential maximum, $|\theta_i-\pi|$,
 is determined solely by the BD distribution during inflation and given by 
 \beq
\ab{\theta_i-\pi}\simeq 
2.4\times 10^{-9}   \left(\frac{N_*}{30}\right) \(\frac{\x^4}{n_{\rm inf}^2-1} \)^{1/9} 
 \(\frac{H_{\rm inf}}{H_{\rm inf}^{\rm T}}\)^2~~~[ {\rm for ~odd~}n_{\rm inf}].
 \eeq
 This is same as Eq.~\eqref{result} with taking $p=2.$

\subsection{Experimental and observational implications}
Here we discuss implications of our $\pi$\hspace{-0.2mm}nflation scenario 
for various experiments and observations. 

\subsubsection{QCD axion DM}
In our $\pi$\hspace{-0.2mm}nflation model (\ref{eq:DIV}), the initial condition of the QCD axion is set to be
near the potential maximum as in Eq.~(\ref{result}), which determines the abundance of the QCD axion
as a function of the inflation parameters.
 Then, assuming that the QCD axion explains
DM, we can relate the QCD axion mass, $m_a$, to the $\pi$\hspace{-0.2mm}nflaton mass, $m_\phi$.

\begin{figure}[!t]
\begin{center}  
     \includegraphics[width=105mm]{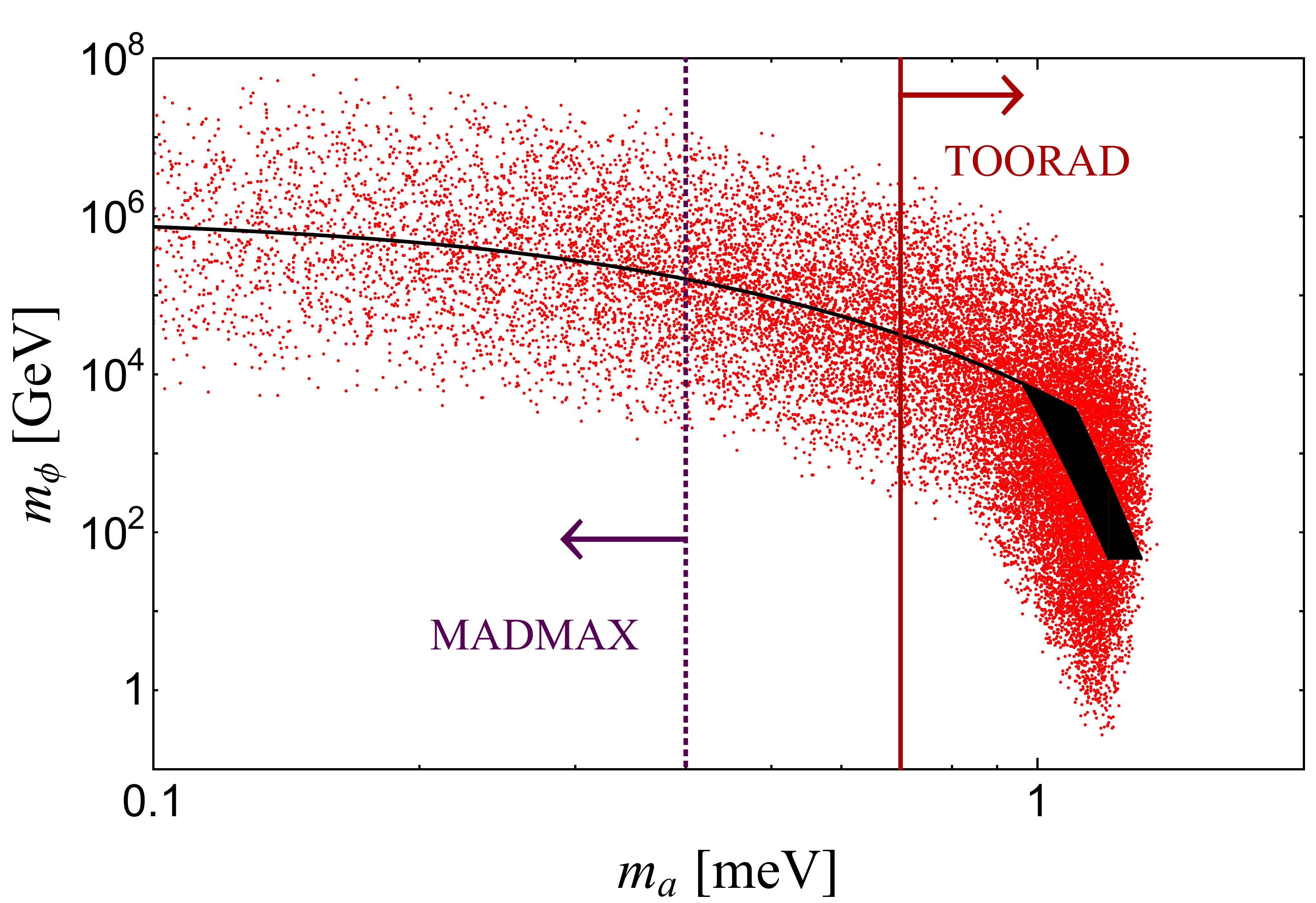}
      \end{center}
\caption{
The relation between the QCD axion mass $m_a$ and the inflaton ($\pi$\hspace{-0.2mm}nflaton) mass $m_\phi$,
where the QCD axion is assumed to explain all DM.
The black line is based on the model  (\ref{eq:DIV}) with $n_{\rm inf}=2$. 
The width of the black line becomes broader at $m_a \gtrsim 1$\,meV, where
the deviation from the hilltop is determined by the inflaton dynamics rather than the Bunch-Davies
distribution, and we vary $\xi$ in the range of (\ref{range}) to include the uncertainties 
 of the inflaton parameters. Note that $m_\phi$ as well as $H_{\rm inf}$ decrease as $m_a$ increases.
We also show the sensitivity reach of MADMAX, and TOORAD. 
}
\label{fig:res1} 
\end{figure}

In Fig.\,\ref{fig:res1}, we show the relation between $m_a$ and $m_\phi$ by the black solid line
based on the $\pi$\hspace{-0.2mm}nflation model (\ref{eq:DIV}) with $\xi=[0.001,1]$, 
 assuming the QCD axion DM, $\Omega_a h^2=0.12$.  
The red points around it are generated by varying  the potential height (\ref{eq:V0}),
the quartic coupling (\ref{eq:ratio}),  and the inflaton mass (\ref{mphi})
 by a factor of order unity. They show how an extension of the inflation model
changes the result, and more details will be given below. 
We also show the projected sensitivity reaches of the MADMAX experiment (left to the purple dotted line)
and the TOORAD experiment (right to the red solid line), both of which look for the QCD axion DM through
its coupling to photons. Here we have assumed the axion-photon coupling for the KSVZ axion.
The $\pi$\hspace{-0.2mm}nflaton $\phi$ can also be searched for in beam dump experiments
if the mass is small enough. The region with inflaton mass below $\O(1)\GEV$ may be searched for 
in the SHiP experiment~\cite{Anelli:2015pba,Alekhin:2015byh,Dobrich:2015jyk, Dobrich:2019dxc} with certain couplings of $\f$ to the SM particles.
The right boundary is set by the condition \eq{cond1}.
In the entire parameter region shown here, the isocurvature perturbations are well below the current limit. 
See also Fig.~\ref{fig:res3}.

So far we have studied the minimal multi-natural inflation in which
the inflaton potential consists of the two cosine terms, but one can extend it to a model
with several cosine terms contributing to the potential,
\begin{eqnarray}
\label{geneH}
V_{\rm inf}(\phi) \,=\,  \sum_{n}{\Lambda^4 \frac{ \k_n}{n^2}\cos\left(n \frac{\f}{f_\f}+\Theta_n \right)}+{\rm const}. ~.
\end{eqnarray}
Here $\k_n, \AND \Theta_n$ are relative height and phase for mode $n$, respectively.
As in the minimal case, we require that the potential is extremely flat around the potential maximum 
$\phi \approx 0$ due to the cancellation among those terms. If there is no extra cancellation or
fine-tuning, successful inflation is still possible without significantly changing various relations between
the parameters from the minimal case. In particular, we assume $\Theta_n=\O\((f_\f/M_{\rm pl})^3\)$,
$|\d|= \({2\x}\)^{1/3} {f_\f}/{M_{\rm pl}}$ with $\x=\O(10^{-3} -1)$,
 and there is no local minimum between $\phi_{\rm inf} \approx 0$ and $\phi_{\rm min} \approx \pi f_\f$.
On the dimensional grounds, the CMB normalization similarly fixes the quartic coupling as
$\lambda= \O(\L^4/f^4) =\O(10^{-13})$,  the inflaton mass at the potential minimum
is $m_\f= \O(\L^2/f)$ unless all $n$ are odd, and the inflation energy density is $V_0=\O(\L^4).$ 
To be concrete, we parameterize them as 
\begin{align}
m_\f & = C_1 \sqrt{2}\frac{\L^2}{f_\f}, \laq{inf1}\\
\frac{\L}{f_\f} &= C_2  \cdot 10^{-3} \(\frac{30}{N_*}\) \\
 V_0 & = C_3\, \L^4, \laq{inf3}
 \end{align}
where $C_i$ are constants of $\O(1)$. The red points in Fig.\,\ref{fig:res1} are generated by 
randomly taking $C_i=[0.1,10]$.

\subsubsection{$\Delta N_{\rm eff}$}
Another important prediction of our scenario is that the QCD axions are thermally populated after reheating, contributing 
to the effective neutrino species, $\D N_{\rm eff}$.
The $\pi$\hspace{-0.2mm}nflation is low-scale inflation, and the decay constant $f_\phi$ is smaller than
$\O(10^{12})\GEV$. The couplings of $\phi$ to the SM particles suppressed by $f_\phi$ often
lead to the instantaneous reheating due to the perturbative decay and dissipation effects.
Then, the reheating temperature is given by
\beq T_R\simeq\({\frac{90}{\pi^2g_*}}\)^{1/4}\sqrt{ H_{\rm inf}M_{\rm pl}}.\eeq
This is the case especially if $\phi$ has a coupling to the top quark, and the reheating temperature can be
as high as  $T_R=\O(10^{7-8})\GEV$ for $m_\f =\O(10^{5-6})\GEV$~\cite{Daido:2017tbr, Takahashi:2019qmh}.

\begin{figure}[!t]
\begin{center}  
     \includegraphics[width=105mm]{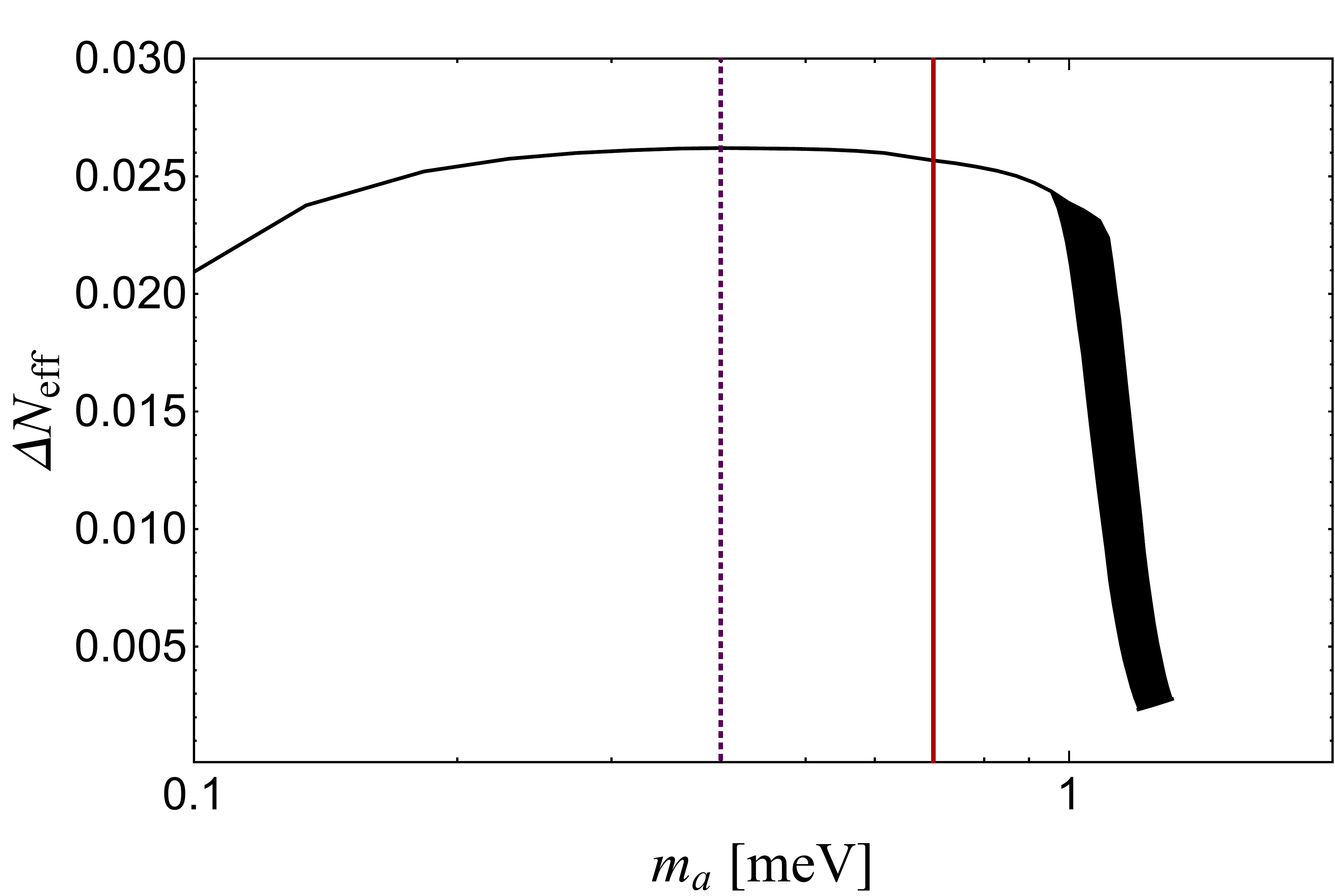}
      \end{center}
\caption{The contribution of the QCD axion to effective neutrino species as a function of
 $m_a$ based on the model  (\ref{eq:DIV}) with $n_{\rm inf}=2$, 
 where the instantaneous reheating is assumed.  The width of the black band becomes broader
 at $m_a \gtrsim 1$\,meV corresponding to the uncertainties of $\xi$ (cf. Eq.~(\ref{range})). The predicted $\Delta N_{\rm eff}$
 also sharply drops at $m_a \gtrsim 1$\,meV because the required inflation scale is so low that axions are
 not thermalized. The slight decrease at the lower end of the mass is due to the increase of $f_a$ relative
 to the reheating temperature. 
 The vertical dotted and solid lines represent sensitivity reach of MADMAX and TOORAD, respectively. }\label{fig:res2} 
\end{figure}

Note that the PQ symmetry is not restored after inflation for the parameters of our interest, because the
$H_{\rm inf}$ is bounded above, $H_{\rm inf} \lesssim 1$\,GeV, for the QCD axion to acquire
its potential during inflation. 
Interestingly though, since the decay constant of the QCD axion is not many orders of magnitude 
larger than the reheating temperature for most of the parameter region, the QCD axions can be 
produced from the thermal scattering via the gluon coupling.
Using the numerical results of  Ref.\,\cite{Salvio:2013iaa}, 
we estimate $\D N_{\rm eff}$ for $n_{\rm inf}=2$  by assuming instantaneous reheating, and 
the result is shown  in Fig.\,\ref{fig:res2}. 
$\D N_{\rm eff}$ can be as large as $0.02-0.026$ for $m_a = 0.1 - 1$\,meV, which can be tested
 by future CMB and BAO observations~\cite{Kogut:2011xw, Abazajian:2016yjj, Baumann:2017lmt}.

\subsubsection{Axion isocurvature perturbations}
\begin{figure}[!t]
\begin{center}  
     \includegraphics
[width=105mm]{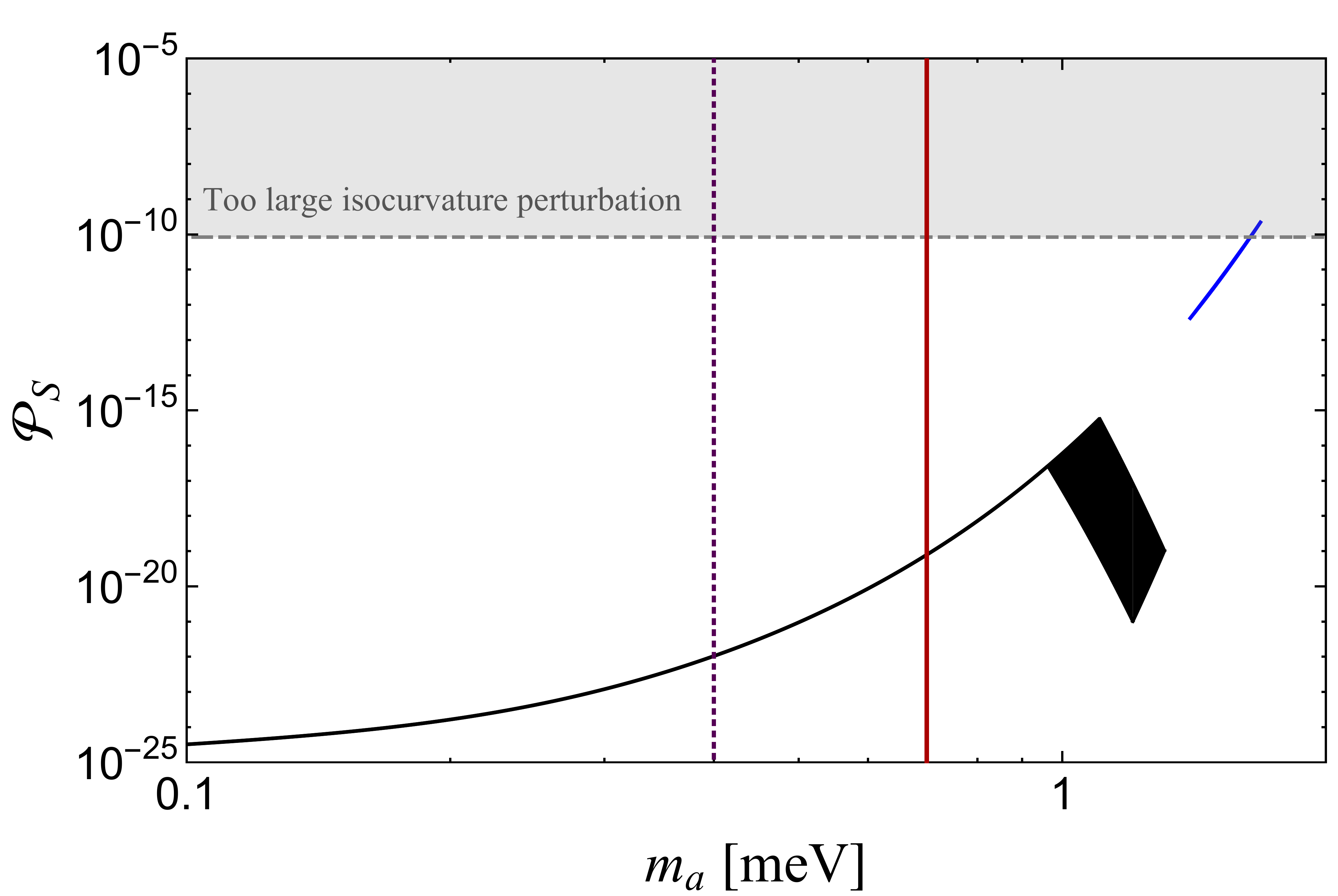}
      \end{center}
\caption{The axion isocurvature perturbation predicted by the $\pi$\hspace{-0.2mm}nflation 
model  (\ref{eq:DIV}) with  $n_{\rm inf} = 2$ (black) and $n_{\rm inf} = 3$ (blue).  
The inflation scale decreases as $m_a$ increases (cf. Fig.~\ref{fig:res1}).
The predicted isocurvature increases for $m_a \lesssim 1$\,meV because $\theta_i$ approaches $\pi$, which overcomes
the  decrease of $H_{\rm inf}$ (see Eqs.~(\ref{Siso}) and (\ref{result})). On the other hand,
the deviation from the hilltop is determined by the inflaton dynamics for $m_a \gtrsim 1$\,meV,
and the predicted inflation scale sharply drops, leading to the suppression of the isocurvature. 
The gray region is excluded by too large isocurvature perturbations. 
The vertical dotted and solid lines represent sensitivity reach of MADMAX and TOORAD, respectively.
}\label{fig:res3} 
\end{figure}

The axion isocurvature perturbations are enhanced in the hilltop limit~\cite{Lyth:1991ub,Kobayashi:2013nva}.
We show in Fig.\,\ref{fig:res3} the predicted isocurvature perturbations of the QCD axion DM 
in the model  (\ref{eq:DIV}) with $n_{\rm inf} = 2$ (black) and with $n_{\rm inf}=3$ (blue).
Note that the predicted mass range is different between the cases of even and odd $n_{\rm inf}$. 
The top gray region is excluded because of too large isocurvature perturbations. We have also estimated
the non-Gaussianity which is also mildly enhanced~\cite{Kobayashi:2013nva}, but it is well within the
current bound~\cite{Akrami:2019izv}.

In the case of $n_{\rm inf}=2$,  the isocurvature perturbation becomes the largest around $m_a \simeq 1$\,meV.
This is because of the dependence of ${\cal P}_{S}$ on $H_{\rm inf}$ (see Eq.~(\ref{isoeven})). As one can see from the figure, it is well below the current (and future) bound.

In the case of $n_{\rm inf}=3$, we have $\delta = 0$ (cf. Eq.~(\ref{d0})) due to the upside-down symmetry.
Also, the inflaton mass $m_\phi$ is related with the curvature at the potential maximum, 
which is bounded as $|V''_{\rm inf}(\phi_{\rm inf})| \lesssim H^2_{\rm inf}$ to satisfy the slow-roll condition~\cite{Daido:2017wwb, Daido:2017tbr}. In the figure we vary it in the following range,
\begin{align}
   m_{\phi} =[0.1-1] \,H_{\rm inf}.
\label{eq:mphi-f-oddn}
\end{align}
Due to its small mass compared to the case of even $n_{\rm inf}$, the inflaton
is so long-lived that its late-time decay into the SM particles tends to cause
cosmological difficulties, e.g. spoiling the success of the big-bang nucleosynthesis. 
If the mass is sufficiently small, however, the inflaton becomes stable on a cosmological time scale. 
Even if the inflaton mass is small, the reheating proceeds through a combination of the perturbative 
decay and dissipation effects, and the inflaton condensate can completely evaporate~\cite{Daido:2017tbr}.  
In this case the inflaton particles are also thermalized in the plasma, contributing to hot DM (or dark radiation).
The inflaton mass is constrained by the structure formation as\cite{Osato:2016ixc,Daido:2017tbr}
\beq
\laq{str}
m_\f\lesssim 7.7\EV ~~~[{\rm for~odd}~n_{\rm inf}].
\eeq
This sets the upper bound on $H_{\rm inf}$, which can be translated to the lower bound
on the mass of the QCD axion $m_a$.
The coupling of $\phi$ to gluons is constrained by SN1987~\cite{Mayle:1987as,Raffelt:1987yt,Turner:1987by, Chang:2018rso} due to the duration of the neutrino burst, leading to $f_\f\gtrsim 10^{8}\GEV$. This sets an upper bound on $m_a$. 
In addition we impose the condition \eq{cond1} as well. 
These constraints restrict $H_{\rm inf}$ to be small, and the predicted isocurvature perturbation 
is enhanced to be at a detectable level due to ${\cal P}_{S}\propto H_{\rm inf}^{-2}$. 
The CMB-S4 experiment will be able to improve the bound on the isocurvature perturbation  by a factor of five compared to {\it Planck}~\cite{Abazajian:2016yjj}.
Therefore, the scenario with odd $n_{\rm inf}$ can be probed by
 searching for the QCD axion DM in the TOORAD experiment and the isocurvature perturbation in the future
 CMB observations.\footnote{
 Note that we have here assumed that the QCD axion is the dominant DM. 
 When $n_{\rm inf}$ is odd, it is possible for $\phi$ to be (a part of) DM if the reheating is incomplete~\cite{Daido:2017wwb, Daido:2017tbr}. In this case, the experimental signal of the QCD axion DM 
may be suppressed. }

\section{Discussion and conclusions}
In this section we briefly discuss other implications of the $\pi$\hspace{-0.2mm}nflation mechanism.
\paragraph{A comment on spontaneous CP breaking}
In an $\SU(N)$ gauge theory with $\theta\simeq \pi$, under several conditions,
there can be a spontaneous CP breaking where the first derivative of the partition 
function with respect to $\theta$ is non-vanishing~\cite{Witten:1980sp, Smilga:1998dh, Gaiotto:2017yup, Kitano:2017jng}.
It is discussed in Ref.\,\cite{DiVecchia:2017xpu} that in the SM at the zero temperature such breaking does not occur due to the small but non-degenerate up and down quark masses. 
Although at finite temperature a phase transition to the broken phase was argued to be possible, this possibility is currently disfavored by the recent lattice data. 
Note however that a hilltop QCD axion may be ill-defined if the up and down quark masses are degenerate or if
they are much heavier than the effective QCD scale in such an extension that
the Higgs has an expectation value much larger than the weak scale.
In our scenario, the quark masses as well as the QCD scale are the same as in the SM,
and the spontaneous CP breaking is unlikely. 

\paragraph{ALP DM, hilltop inflation, and $N$-flation}
So far we have considered a scenario in which the initial condition of the QCD axion is set near the
potential maximum by the $\pi$\hspace{-0.2mm}nflation. It is straightforward to extend it to an ALP 
which does not solve the strong CP problem. The ALP ($\phi_2$) will have a potential $V_{2}( \f ,\f_2)$ instead 
of the second term in \Eq{tot}, and the initial condition of $\phi_2$ can be set close to the potential
maximum of $V_2$ after inflation. This will generically enhance the abundance of the ALP as well as its isocurvature perturbations. The ALP DM with a hilltop initial condition
is recently studied in Ref.~\cite{talk2}.\footnote{
This actually motivated us to revisit our idea in 2017~\cite{Daido:2017wwb} and study it in detail.
}
The ALP can also be a curvaton~\cite{Enqvist:2001zp,Lyth:2001nq, Moroi:2001ct}, and it has interesting implications such as mild enhancement of the non-Gaussianity in the hilltop limit~\cite{Kawasaki:2008mc,Kawasaki:2011pd,Kawasaki:2012gg}.  Similarly, the $\pi$\hspace{-0.2mm}nflation can explain the initial condition for 
axion hilltop inflation~\cite{Czerny:2014wza,Czerny:2014xja,Czerny:2014qqa,Higaki:2014sja, Croon:2014dma}. 
In general, the hilltop inflation requires a fine-tuning of the initial condition of the inflaton near the potential
maximum. Although such fine-tuning may be compensated by the eternal inflation, it is neat that there
is a dynamical way to put the inflaton near the top of the potential.

The $\pi$\hspace{-0.2mm}nflation requires a mixing between axions, and such mixings may be ubiquitous 
in the axion landscape~\cite{Higaki:2014pja,Higaki:2014mwa}. In particular, if the $\pi$\hspace{-0.2mm}nflaton is mixed with multiple
axions, it is possible that  all of them are placed near their potential maxima after inflation.
The abundances of those axions can be similarly enhanced, and the viable parameter region  to explain DM 
will be different from the usual case of a single ALP DM. Alternatively, those axions
may drive a period of inflation as in the $N$-flation~\cite{Liddle:1998jc,Dimopoulos:2005ac,Kim:2010ud}.
Thus, the $\pi$\hspace{-0.2mm}nflation can provide a plausible initial condition for the $N$-flation.

 \paragraph{Chain of $\pi$\hspace{-0.2mm}nflation}
As mentioned above, the $\pi$\hspace{-0.2mm}nflation can set a plausible initial condition
for the hilltop inflation. This may lead to a chain of $\pi$\hspace{-0.2mm}nflation in a certain
set-up.
One can consider the following simple potential,
\beq
V_{\rm inf}(\phi_i)=\L_1^4\ \cos{\(\frac{\f_1}{f_1}\)} - \sum_{i=1}^{N} \L_{i+1}^4 \cos{\(\frac{\f_{i}}{f_{i}}+\frac{\f_{i+1}}{f_{i+1}}\)} + {\rm const.}
\eeq
where $f_i$ is the decay constant, 
$\phi_i$ respects the discrete shift symmetry, $\phi_i \rightarrow \phi_i + 2 \pi f_i$, 
the dynamical scales, $\L_i$, satisfy $\L_i\gg \L_{i+1}$, and the last constant term is introduced to make
the cosmological constant vanishingly small in the present vacuum.
Without loss of generality we consider the range of $-\pi < \phi_i/f_i \leq \pi$. The potential minimum is
located at $\phi_i/f_i = \pi$. Let us suppose that $\phi_1$ is initially around the origin and
drives inflation. To this end, we take all $f_i$ comparable to or larger than the Planck scale for simplicity,
One may replace the cosine-terms for $\f_i$ with multiple cosine terms or other functions having a flat plateau,
which allow inflation without introducing super-Planckian decay constants. For $\phi_1 = 0$, the potential is minimized at $\phi_2 = \phi_3 = \cdots = \phi_{N+1} = 0$.
 If the inflation driven by $\phi_1$ 
lasts sufficiently long, $\phi_2$ (as well as $\phi_i$ with $i>2$) will follow the BD distribution peaked at the origin.
Then, after the inflation ends, $\phi_1$ will move from the origin to $\pi f_1$. As a result, $\phi_2 = 0$ is
now the potential maximum. Then, $\phi_2$ will drive the next inflation when its potential energy comes to dominate
the Universe. Thus, this chain of $\pi$\hspace{-0.2mm}nflation can continue. Note that the observed CMB density perturbations are generated during the last $50$ (or smaller)  $e$-folds, and so, most part of the inflaton potentials 
are not constrained by observations in this set-up.

 \paragraph{Old-type $\pi$\hspace{-0.2mm}nflation}
 So far we have considered a case in which the $\pi$\hspace{-0.2mm}nflation is realized by a slow-roll inflation.
 However, our definition of the $\pi$\hspace{-0.2mm}nflation is such that its dynamics gives a phase shift close to
 $\pi$ for another axion through mixing, and so, it may be realized by the so-called old-type inflation that
 ends by bubble formation~\cite{Guth:1980zm}. Such a possibility was discussed in Ref.~\cite{Daido:2017wwb}.
Let us suppose that $\phi$ is trapped in a false vacuum at the origin $\phi=0$, and it tunnels to the true vacuum
at $\phi \approx \pi f_\phi$ through bubble nucleation. $\phi_2$ is another axion that has a mixing with $\phi$.
If $\phi_2$ is light, and if the old inflation lasts sufficiently long, the probability distribution of $\phi_2$
is given by the BD distribution peaked at the potential minimum during inflation (say, $\phi_2 = 0$). 
After the tunneling along the $\phi$ direction, the potential for $\phi_2$ will be flipped due to the phase shift
of $\phi$. The $\pi$\hspace{-0.2mm}nflation similarly works in this case.

\paragraph{QCD axion inflation}
So far, we have assumed that the inflaton dynamics is not affected by the potential generated by non-perturbative QCD effects.
In fact, however, successful inflation is possible even if a combination of $\phi$ and $a$ becomes sufficiently heavy due to 
the non-perturbative QCD effects. The inflaton dynamics is similar to the hybrid inflation model~\cite{Copeland:1994vg,Dvali:1994ms,Linde:1997sj}, and it is studied in detail in Ref.~\cite{Daido:2017wwb} in the context of the ALP inflation. 

Let us first assume that a linear combination of $\phi$ and $a$ appearing in the second term of \Eq{tot} acquires
a heavy mass during inflation. We also assume even $n_{\rm inf}$ for which $\phi$ becomes heavy in the present vacuum,
so that $a$ is identified with the QCD axion in the low energy. 
In this case, the light and heavy mass eigenstates at $(\f, a) \approx (0,\pi f_a)$ 
are given by
\begin{align}
\label{A1}
A_L & \equiv \frac{1}{\sqrt{1+k^2}} \left(a - k \f \right),\\
A_H & \equiv \frac{1}{\sqrt{1+k^2}} \left(\f + k a \right),
\label{A2}
\end{align}
where $k \equiv f_\f/ n_{\rm mix} f_a$ is assumed to be much smaller than unity, for our purpose. 
The mass of $A_H$ is
\begin{align}
M_H & = \sqrt{1+k^2} n_{\rm mix} \frac{\sqrt{\chi}}{f_\f}
\end{align}
If $M_H > H_{\rm inf}$, we integrate out $A_H$ by seting $A_H = 0$, and
we can  express the inflaton potential $V_{\rm inf}(\phi)$
in terms of $A_L$ using the relation,
\begin{align}
\label{pAHL}
\phi = \frac{1}{\sqrt{1+k^2} }(A_H - k A_L).
\end{align}
Setting $\Theta = 0$ and $\kappa = 1$ for simplicity, the quartic potential for $\phi$ is now given by
\begin{align}
V_{\rm inf} &= V_0 - \lambda \phi^4 + \cdots \\
                  &= V_0 -  \frac{\l k^4}{(1+k^2)^2}  A_L^4  + \cdots
\end{align}
where we have substituted (\ref{pAHL}) and set $A_H = 0$.
Note that, for $k \ll 1$, the inflaton $A_L$ is almost identical to the QCD axion, $a$. 
It is the QCD axion that drives the slow-roll inflation through the mixing with another axion $\phi$.

 From the CMB normalization (\ref{eq:ratio}), we obtain
\beq
\frac{\L}{f_\f} k\simeq \frac{\L}{ n_{\rm mix} f_a } \simeq 1.2\times 10^{-3 }\(\frac{3}{n_{\rm inf}^2-1}\)^{\frac{1}{4}}\(\frac{30}{ N_*}\)^{\frac{3}{4}}.
\eeq
Thus, one can relate the inflation scale to the QCD axion decay constant,
\beq
\laq{QCDinf}
H_{\rm inf} \simeq  1\times10^{-8}\GEV n_{\rm mix}^2 \(\frac{3}{n_{\rm inf}^2-1}\)^{\frac{1}{2}}
\(\frac{30}{ N_*}\)^{\frac{3}{2}} \( \frac{f_a}{10^{8}\GEV}\)^2.
\eeq
When the $A_L$ reaches the point where the curvature along $A_L$ becomes comparable to $M_H$,
 the inflaton starts to oscillate towards $\f$ direction and $\f$ decays to reheat the universe.
 In this sense, $\f$ is like a waterfall field in hybrid inflation.
 
 The condition $M_H > H_{\rm inf}$ reads
 \begin{align}
 \label{upper}
 f_\phi < 7{\rm\,TeV}\,  n_{\rm mix}^{-1} \(\frac{n_{\rm inf}^2-1}{3}\)^{\frac{1}{2}}
\(\frac{ N_*}{30}\)^{\frac{3}{2}} \( \frac{10^{8}\GEV}{f_a}\)^{2}.
 \end{align}
Another constraint comes from the perturbativity of the inflaton potential $V_{\rm inf}(\f)$,
\beq
\L  \lesssim 2\pi f_\f,
\eeq
which gives
\begin{align}
f_\phi \gtrsim 20 \TEV\,  n_{\rm mix} \(\frac{3}{n_{\rm inf}^2-1}\)^{\frac{1}{4}}\(\frac{30}{ N_*}\)^{\frac{3}{4}}.
\end{align}
This can be consistent with (\ref{upper}) when $n_{\rm inf}\geq 4$ or by extending the inflation model as \Eqs{inf1}-\eq{inf3}, and we are left with parameters \beq f_\phi \sim 10\TEV \AND f_a \sim 10^8\GEV.\eeq
The parameters as well as the inflation scale \eq{QCDinf} are predictions of the QCD axion inflation.

\paragraph{A possible connection with non-linear sigma models}
The $\pi$\hspace{-0.2mm}nflation is driven by an axion or a pseudo Nambu-Goldstone (NG) boson,
which may be identified with one of those appearing in non-linear sigma (NLS) models.
In particular, SUSY NLS models with a compact K\"{a}hler manifold are interesting as they 
can explain the origin of the three families where the leptons and quarks are in NG multiplets~\cite{Kugo:1983ai, Yanagida:1985jc}. 
A consistent theory coupled to gravity predicts an axion-like multiplet, $S$, and/or NG bosons of $\U(1)$ symmetries~\cite{Komargodski:2010rb, Kugo:2010fs}.  
See Ref.~\cite{Yanagida:2019evh,Yamaguchi:2016oqz,Yin:2016shg,Yanagida:2018eho,Endo:2019bcj} 
for other roles of $S$ and its phenomenological implications. Although the imaginary part of $S$ is not 
coupled to gluons in the minimal set-up, the NG bosons may be identified with the QCD axion and
$\pi$\hspace{-0.2mm}nflaton.

\subsection*{Conclusions}
In this paper we have shown that the initial angle of the QCD axion $a$ can be naturally close to $\pi$,
if it has a mixing with another axion $\phi$ that induces a phase shift close to $\pi$. We have studied the case
in which $\phi$ drives inflation, and we call it $\pi$\hspace{-0.2mm}nflation. \
If the Hubble parameter during 
inflation is below the QCD scale, if the inflation lasts sufficiently long,  the QCD axion follows the BD distribution 
peaked at $\pi$. Interestingly, although the initial angle of the QCD axion is close to $\pi$,  we generically expect
a small deviation from it  due to either the BD distribution or the $\pi$\hspace{-0.2mm}nflaton dynamics (see \Eq{dtheta}
or (\ref{delta})), thereby avoiding the domain-wall problem. 

 The $\pi$\hspace{-0.2mm}nflation  enables the QCD axion to explain DM by the misalignment mechanism 
 even if its  decay constant is smaller than conventionally considered. Note that we do not have to introduce any  explicit PQ breaking terms in contrast to the production mechanism
using the collapse of the string-wall network~\cite{Kawasaki:2014sqa,Ringwald:2015dsf}.  Specifically, our mechanism works for
 $f_a \gtrsim 2.9 \times 10^9\GEV$ or equivalently, $m_{a} \lesssim 2.0\times 10^{-3}\EV$. The QCD axion DM 
 in this mass range can be tested by e.g. ORGAN, MADMAX,  TOORAD and IAXO experiments. We have also shown that 
 the $\pi$\hspace{-0.2mm}nflaton can be searched for at the SHiP experiment in a corner of the parameter space,
 and discussed implications for $\Delta N_{\rm eff}$ and axion isocurvature perturbations. 

Lastly, let us emphasize that the low-scale inflation satisfying $H_{\rm inf} \lesssim 1$\,GeV and $N \gg H_{\rm inf}^2/m_a^2$
enables the QCD axion to explain DM  for both $f_a \gg 10^{12}$\,GeV \cite{Graham:2018jyp,Guth:2018hsa,Ho:2019ayl} and $f_a \ll 10^{12}$\,GeV. In the former, $\theta_i \ll 1$
is realized by the BD distribution around $\theta_i = 0$, while in the latter, $\theta_i \approx \pi$ is realized by the 
$\pi$\hspace{-0.2mm}nflation mechanism.

\section*{Acknowledgments}
F.T. thanks K. Harigaya for discussion on the axion dynamics, T. Sekiguchi for clarification of the current 
limit on non-Gaussianity of the isocurvature perturbation, and the organizers and speakers 
of the CERN-Korea TH Institute for
hospitality and lively discussion, where the present work was initiated. W.Y. thanks R. Kitano for useful discussion on the spontaneous CP breaking. 
This work is supported by JSPS KAKENHI Grant Numbers
JP15H05889 (F.T.), JP15K21733 (F.T.),  JP17H02875 (F.T.), 
JP17H02878 (F.T.),  and by World Premier International Research Center Initiative (WPI Initiative), MEXT, Japan, and by 
NRF Strategic Research Program NRF-2017R1E1A1A01072736 (W.Y.).

\end{document}